\documentclass[sn-basic]{sn-jnl}

\usepackage{graphicx}%
\usepackage{multirow}%
\usepackage{amsmath,amssymb,amsfonts}%
\usepackage{amsthm}%
\usepackage{mathrsfs}%
\usepackage[title]{appendix}%
\usepackage{xcolor}%
\usepackage{textcomp}%
\usepackage{manyfoot}%
\usepackage{booktabs}%
\usepackage{algorithm}%
\usepackage{algorithmicx}%
\usepackage{algpseudocode}%
\usepackage{listings}%
\usepackage{todonotes}

\theoremstyle{thmstyleone}%

\theoremstyle{thmstyletwo}%

\theoremstyle{thmstylethree}%

\usepackage[capitalize,noabbrev]{cleveref}
\usepackage{hyperref}
\hypersetup{colorlinks,linkcolor={blue},citecolor={magenta},urlcolor={blue}}
\usepackage{gensymb}
\usepackage{tcolorbox}
\usepackage{array}
\usepackage{subcaption}
\usepackage{booktabs}       
\usepackage{natbib}

\begin{document}

\title[Learning and Controlling Silicon Dopant Transitions in Graphene using Scanning Transmission Electron Microscopy]{Learning and Controlling Silicon Dopant Transitions in Graphene using Scanning Transmission Electron Microscopy}

\author*[1]{\fnm{Max} \sur{Schwarzer}}\email{maxa.schwarzer@gmail.com}
\equalcont{These authors contributed equally to this work.}

\author*[2]{\fnm{Jesse} \sur{Farebrother}}\email{jessefarebro@gmail.com}
\equalcont{These authors contributed equally to this work.}

\author[3]{\fnm{Joshua} \sur{Greaves}}\email{joshua.greaves@gmail.com}
\author[5]{\fnm{Ekin Dogus} \sur{Cubuk}}\email{cubuk@google.com}
\author[5]{\fnm{Rishabh} \sur{Agarwal}}\email{rishabhagarwal@google.com}
\author[1]{\fnm{Aaron} \sur{Courville}}\email{courvila@mila.quebec}
\author[2,3]{\fnm{Marc G.} \sur{Bellemare}}\email{bellemam@mila.quebec}
\author*[6]{\fnm{Sergei} \sur{Kalinin}}\email{sergei2@utk.edu}
\author*[5]{\fnm{Igor} \sur{Mordatch}}\email{imordatch@google.com}
\author*[1,5]{\fnm{Pablo Samuel} \sur{Castro}}\email{psc@google.com}
\author*[4]{\fnm{Kevin M.} \sur{Roccapriore}}\email{roccapriorkm@ornl.gov}

\affil[1]{\orgdiv{Université de Montréal}, \orgname{Mila}}
\affil[2]{\orgdiv{McGill University}, \orgname{Mila}}

\affil[3]{\orgdiv{Reliant AI}}

\affil[4]{\orgdiv{Oak Ridge National Laboratory}} 
\affil[5]{\orgdiv{Google DeepMind}}
\affil[6]{\orgdiv{Department of Materials Science and Engineering, University of Tennessee, Knoxville, TN 37920}}

\abstract{We introduce a machine learning approach to determine the transition dynamics of silicon atoms on a single layer of carbon atoms, when stimulated by the electron beam of a scanning transmission electron microscope (STEM). Our method is data-centric, leveraging data collected on a STEM. The data samples are processed and filtered to produce symbolic representations, which we use to train a neural network to predict transition probabilities. These learned transition dynamics are then leveraged to guide a single silicon atom throughout the lattice to pre-determined target destinations. We present empirical analyses that demonstrate the efficacy and generality of our approach.}

\keywords{Microscopy, Machine Learning}

\maketitle

\section{Introduction}

Sub-atomically focused electron beams in scanning transmission electron microscopes (STEMs) can induce a broad spectrum of chemical changes, including defect formation, reconfiguration of chemical bonds, and dopant insertion. Several groups have shown the feasibility of direct atomic manipulation via electron beam stimulation, which holds great promise for a number of downstream applications such as material design, solid-state quantum computers, and others \citep{jesse18direct,susi17manipulating,dyck17placing,tripathi18electron,dyck18building}. One of the challenges for advances in this space is that these types of atomic manipulation rely on manual control by highly-trained experts, which is expensive and slow.

The ability to accurately automate this type of beam control could thereby result in tremendous impact on the feasibility of atomic manipulation for real use cases. A critical requirement for this automation is accurate estimation of the transition dynamics of atoms when stimulated by focused electron beams. To date, the microscopy community has relied on heuristic estimates for these transition dynamics, with anecdotal evidence of their accurateness. Indeed, the common practice has been to use the physically intuitive, but heuristic, assumption that the optimal beam position is directly on a neighboring atom.

In this paper we present a technique for estimating these atomic transition dynamics using machine learning techniques on collected observations. Our approach consists in a sequence of steps ultimately resulting in a probability distribution over possible beam positions relative to a particular atom, conditioned on the atom's prior position as well as the electron beam's location and dwell time. We demonstrate the effectiveness of our approach by using it to estimate the transition dynamics of a silicon atom on a carbon lattice (graphene).

As evidence that this process can improve beam control, we use the estimated transition probabilities to automate the sequential control of a silicon atom on graphene towards a pre-specified target position. Modern STEMs are capable of this type of automation, and our work paves the way for future advances in automated atomic control.

\section{Related work}
Atomic manipulation was first demonstrated by \citet{stroscio91atomic} by using the tip of a scanning tunneling microscope (STM) to position individual Xenon atoms on the surface of a single crystal surface to form the IBM company logo. 
Further demonstrations, such as quantum corrals and molecular cascades, have demonstrated the potential of the method. Perhaps the application that has attracted the most interest is in using tip-induced atomic motion as an enabling tool for the fabrication of P and other atoms in Si qubits, the building blocks for quantum computers. 

Despite the feasibility of manual control with STM tips, this type of atomic manipulation is limited to metallic/conducting surfaces. On the other hand, while STEMs can manipulate atoms embedded within a several layer thick specimen, it is still a rather haphazard and unpredictable process relative to using STM tips. To date, electron beam induced effects (with a STEM) have been studied purely by human operation, most typically by scanning a raster pattern (where the electron dose tends to be concentrated non-uniformly on one side of the image) in a selected field of view. The more sophisticated experiments involve manual positioning of the electron beam by a human, but this kind of motion is unpredictable and unreliable, and useful statistics are challenging (if not impossible) to glean from experiments conducted in this manner. We note that other experiments have been performed which control the electron beam in non-standard trajectories, effectively performing direct-write beam patterning processes – but with the critical point that the atomic landscape (i.e., position of atoms) is not considered \citep{dyck23atom,dyck23platform,dyck23topDown}.

The potential of the electron beams of scanning transmission electron microscopes to affect matter on the atomic level has been recognized since the early days of the technique. Most of these effects have been generally classified as a beam damage, denoting unwanted changes in materials structure induced by the beam. Indeed, minimization of beam damage, along with the need to increase spatial and energy resolutions, remains one of the three primary drivers behind STEM development, having spurred the high-voltage machines of the 1980s and 1990s and the aberration corrected low-voltage machines of the last two decades. It was also discovered that electron beam effects can be far more subtle, including crystallization and amorphization of oxides and semiconductors \citep{lulli93comparison, yang97low, jencic95electron, robertson96regrowth, frantz01mechanism}.

The emergence of aberration-corrected STEMs have made the atomic-resolution imaging relatively routine, and spurred a new wave of electron beam matter manipulation on the atomic level. The electron beam was shown to be able to deposit single atoms from chemisorbed species \citep{vanDorp12molecule} and form ordered vacancy arrays \citep{jang17inSitu}. Similarly, electron beams have been shown to induce direct atomic motion and creation of functional defects \citep{cretu12electron,yang14direct,susi14silicon}.

The combination of simple beam control and feedback systems has enabled the direct assembly of crystalline materials with a single unit plane precision via directed crystallization and amorphization \citep{jesse15atomic}. These systems have also demonstrated potential for direct single atom dopant movement \citep{jesse18direct}, and finally, the controlled manipulation of Bismuth dopants in bulk silicon \citep{hudak18directed}. In 2016, it was proposed that the combination of machine learning with electron beam manipulation can become a third paradigm for direct atomic construction \citep{kalinin16fire}. In 2017, \citet{dyck17placing}, \citet{susi17towards} and \citet{susi17manipulating} demonstrated single atom manipulation and insertion experiments for silicon in
graphene \citep{susi17manipulating,dyck17placing,tripathi18electron,dyck18building}, an approach soon extended to direct atomic assembly of homo- \citep{dyck18building} and
hetero-atomic artificial molecules \citep{dyck19atom}.

A number of theories for beam manipulation have been proposed, including those based on phonon-assisted knock-on and electronic excitations. However the causal relationship of the electron beam position relative to the silicon atom has only been suggested and demonstrated anecdotally. Physical intuition dictates that the damage mechanism is primarily through momentum transfer or so-called “knock-on” processes; therefore, the ideal placement of the electron beam would seem to be positioned exactly centered on a carbon (first) neighbor. While this is intuitive and appears to have been a successful route by multiple groups, damage mechanisms tend to be complex and are dictated by more than one process. For example, ionization or sputtering processes may be occurring as well, meaning it is unclear if the suggested beam position is actually the ideal one for inducing the most efficient transition of a silicon hop. Moreover, the anecdotal but physically intuitive rule of placing the electron beam on the center of a carbon neighbor is mostly valid only for a direct Si substitution (i.e., 3-fold coordinated silicon). For any other configuration, the rules are already not the same, and the optimal beam position for causing a transition event is not clear.

\section{Problem description}
\label{sec:problemDescription}

\begin{figure}[!t]
\centering
\includegraphics[width=\linewidth]{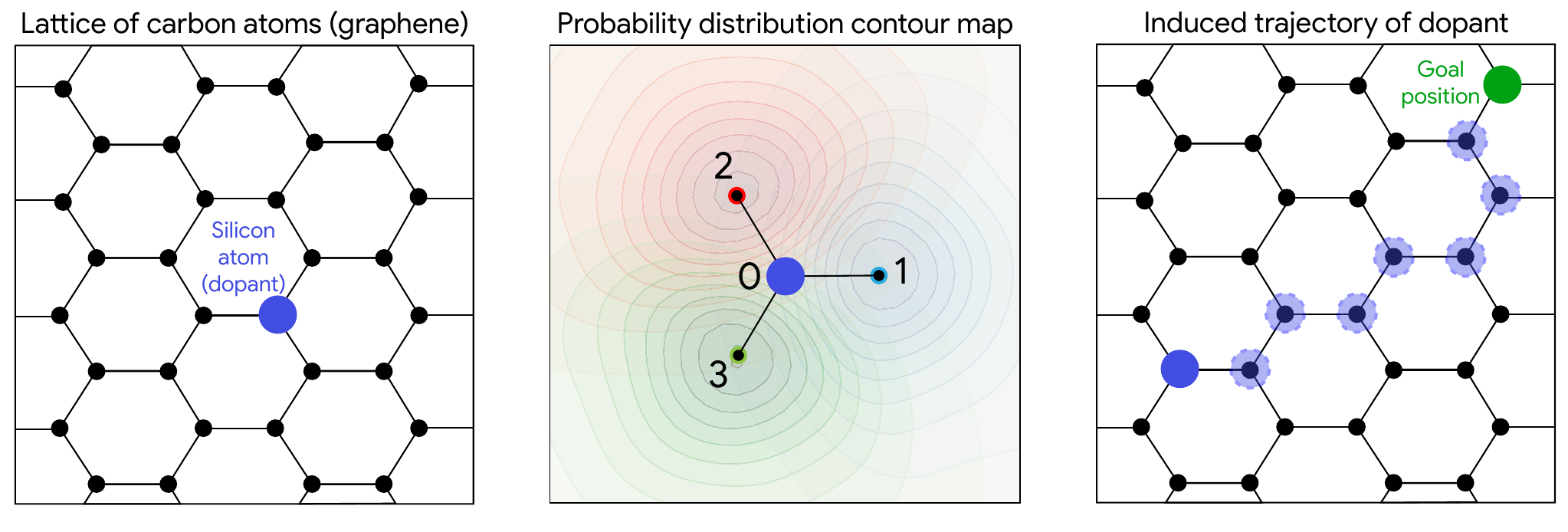}
\caption{{\bf Left:} Illustration of pristine graphene with a single dopant. Carbon atoms depicted with black circles, while the silicon atom in blue; {\bf Center:} Contour map learned by our method, depicting the probability of transitioning to each of the neighbours for different beam positions. The distributions for each of the neighbours are differentiated using three colours, and the numbers indicate neighbour ordering as discussed in \cref{sec:dataFiltering}; {\bf Right:} Example trajectory of dopant towards a goal position.}
\label{fig:problemDescription}
\end{figure}

Our system consists of {\em graphene}: a single layer of carbon atoms arranged in 3-fold configuration (i.e. every carbon atom is connected to three other carbon atoms). On this lattice, a single silicon atom (hereafter referred to as {\em the dopant}) has taken the place of one of the carbon atoms. We focus an electron beam on a position in the area spanned by the dopant and its three carbon neighbours for a specified amount of time (referred to as the {\em dwell time}). This electron beam stimuli can result in the dopant moving to one of its neighbours (by trading places with the respective carbon atom), or in the configuration remaining unchanged\footnote{Note that for long dwell-times this "unchanged" outcome can be a result of the dopant moving twice: once to its neighbour and then back to its original position.}. This configuration has been extensively explored, and hence is an ideal system for exploring this type of automation \citep{wang2014direct,dyck17placing,markevich20uncovering,markevich21mechanism}. We provide an illustration of this configuration in the left panel of \cref{fig:problemDescription}.

Our objective is to learn a probability distribution over the position of the dopant, conditioned on its current position, beam location, and beam dwell time. If accurate, we can use this distribution to determine the optimal beam location and dwell time so as to induce the dopant to move to one of its neighbouring positions. In the center panel of \cref{fig:problemDescription} we can see a heat map depicting the transition probabilities for each of the dopant's neighbours (differentiated with colors) for varying beam locations.

Equipped with these probability maps, we can repeatedly induce transitions of the dopant to neighbouring atoms, resulting in a full trajectory. In other words, we can use a fully greedy strategy to move the dopant to any pre-specified target position on the lattice via a simple shortest-distance path. In the right panel of \cref{fig:problemDescription} we depict one such possible trajectory.

\section{Description of Methodology}
\label{sec:methodology}

\subsection{Data collection}
\label{sec:dataCollection}
Our methodology relies on real data collected with a STEM device, so it is important that the data gathered is informative for the task at hand. Since we are concerned with the transition dynamics of the dopant, our data collection approach is as follows:
\begin{enumerate}
    \item Acquire an initial image of the graphene. For this, we used a field of view of 3 nm with the dopant in the center. A sample image is shown on the bottom-left panel of \cref{fig:methodology}.
    \item Sample a position uniformly within a 2.84 \r{A} radius\footnote{Chosen to be the cumulative length of two carbon-carbon bonds.} of the atom. In the top-center panel of \cref{fig:methodology} we display the normalized beam positions. 
    \item Focus the electron beam at that position for a dwell time drawn from a distribution typically sampled from 1 to 10 seconds. The top-left corner of \cref{fig:methodology} displays the distribution of dwell times.
    \item Acquire a final image of the graphene.
\end{enumerate} 

\begin{figure}[!t]
\centering
\includegraphics[width=\linewidth]{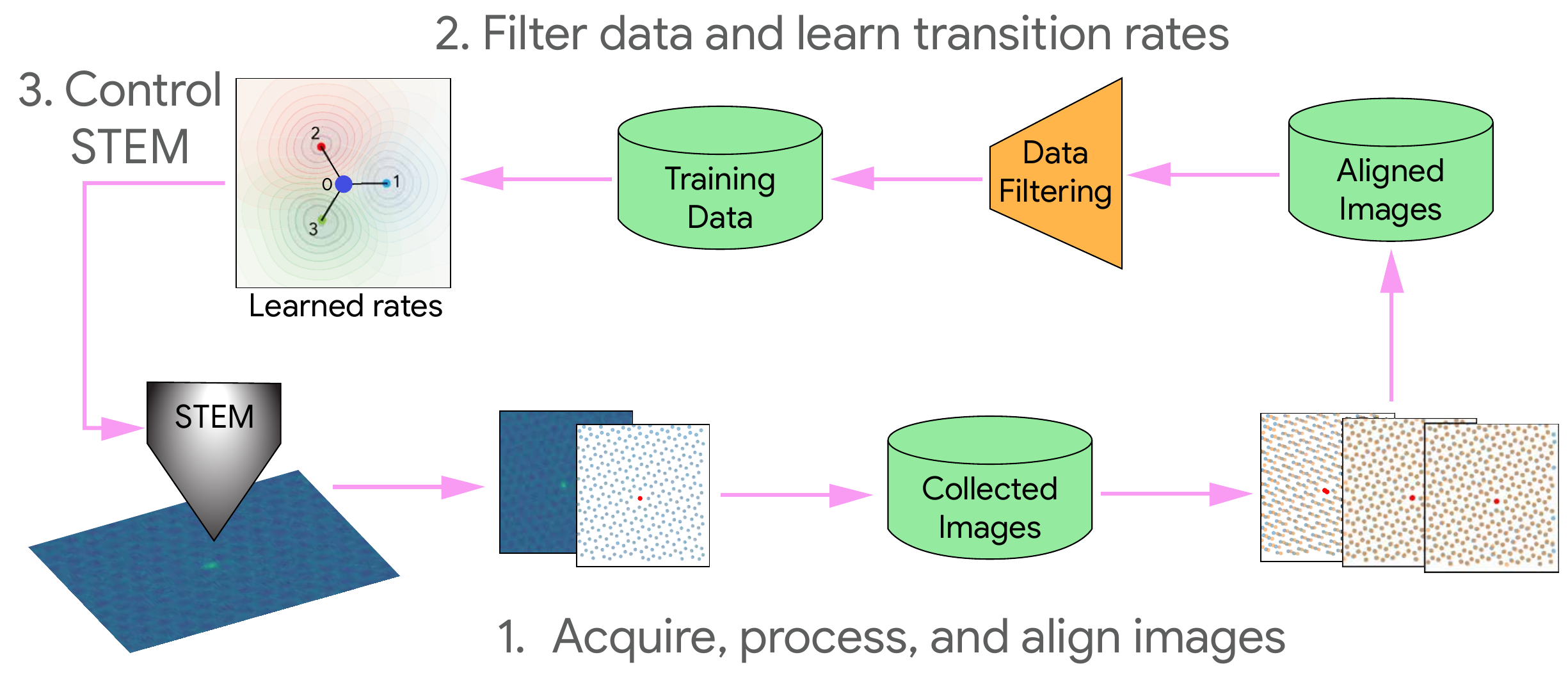}
\caption{An overview of the full pipeline for learning the transition probabilities.}
\label{fig:methodOverview}
\end{figure}
In this section we detail the methodology we used to learn the transition dynamics of electron-beam induced atomic manipulation, illustrated in \cref{fig:methodOverview}. We used a Nion UltraSTEM 100, which grants us access to nearly every microscope control via a Python API. The silicon dopant atoms have been inserted into the lattice in a previous step at a higher (100 kV) accelerating voltage, which is described elsewhere~\citep{roccapriore2023discovering}.

There are a few considerations that are worth mentioning. First, it is important to gather multiple samples using the same beam position and dwell time, as transitions are probabilistic. Second, image acquisition is done using the same electron beam, which implies that imaging itself can cause a transition; to mitigate this, the imaging electron dose should be minimized. Third, placing the electron beam in a known position relative to the silicon should be conducted in a controlled and automated fashion – for this, atomic coordinates must be known with high confidence in as close to real time as possible, and flexible control of the beam position is needed. This last consideration is addressed by using deep ensemble neural networks to identify the atomic species and positions both quickly and reliably~\citep{Ghosh2021}.

\subsection{Atomic alignment}
\label{sec:aligner}

\begin{figure}[!t]
\includegraphics[width=0.95\textwidth]{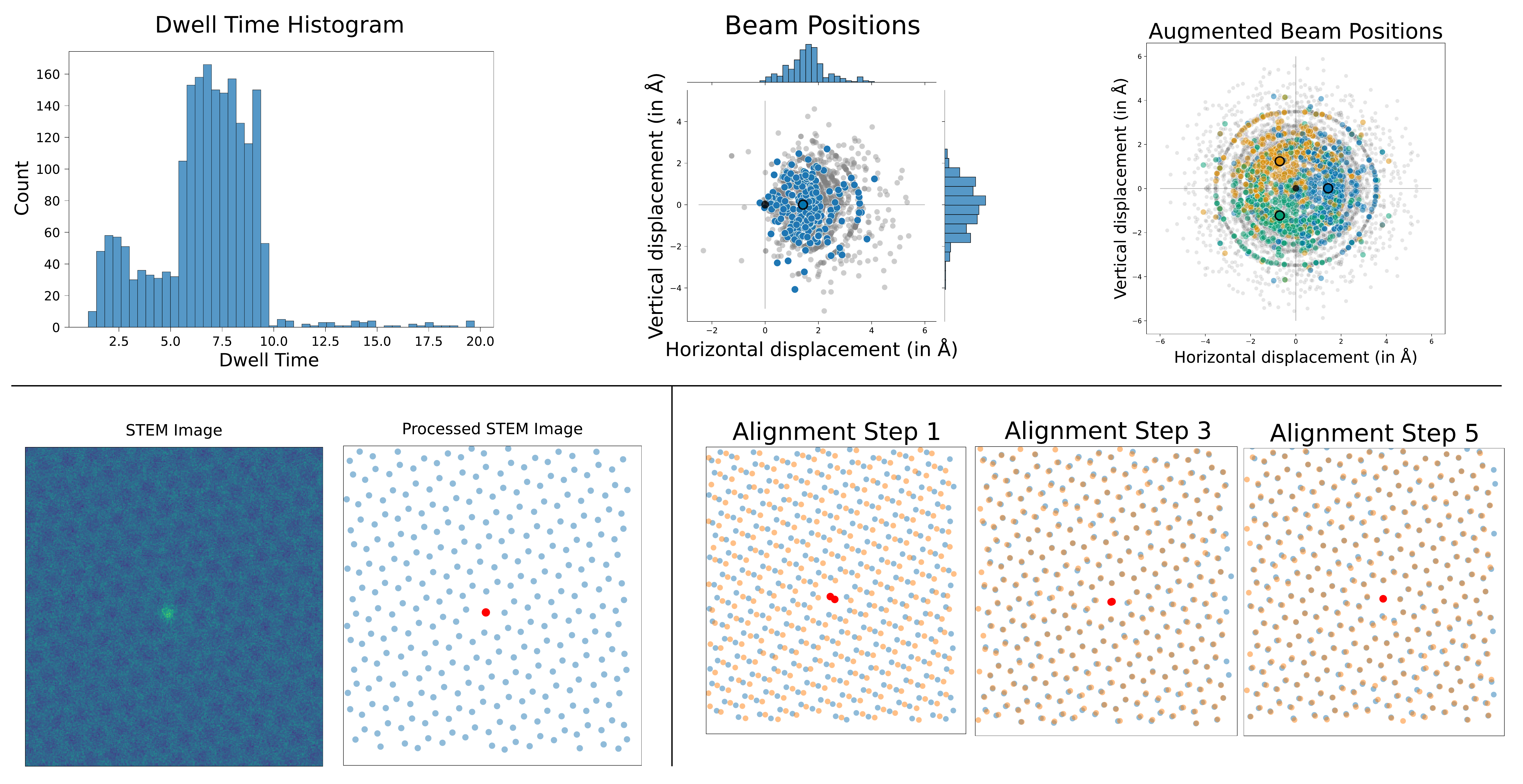}
\caption{{\bf Top row:} Histogram of dwell times (left) and standardized beam positions (center) used for data acquisition; beam position data augmented for three neighbours (right). In the beam position plots, grey and colored circles represent negative and positive transitions, respectively. The colored circles with a black border represent the neighbouring atoms. {\bf Bottom left:} Raw image acquired from the STEM (left), and processed image after atom detection (right). {\bf Bottom right:} The result of conducting multiple alignment iterations with our trained convolutional aligner. Orange dots reflect estimated atom positions in the previous scan, blue dots reflect atom positions in the current scan, and red dots are dopant positions. As can be observed, the discrepancies diminish with additional iterations.}
\label{fig:methodology}
\end{figure}

In the bottom-left panel of \cref{fig:methodology} we display one of the raw images captured with the STEM device, and to the right of it the processed output. The processed image is an ``idealized'' configuration as illustrated in \cref{fig:problemDescription}, which are easier to operate on. While it is relatively simple to identify the atoms in a single image, there are some challenges that arise when using more than one raw image, as discussed in \cref{sec:dataCollection}: the graphene sheet may have physically moved between image acquisition steps (specimen drift), and there may be aberrations (such as warping) caused by the electron beam. To be able to use the images acquired for learning transition dynamics, we need not only to detect the atoms in each of the separate images, but also {\em map} each of the atoms from one image to the next.

We considered three solutions to this problem. The classical approach is to take the cross-correlation between the two scans and estimate the drift to be the arg max of this cross-correlation. Unfortunately, as images of this system are generally dominated by the very bright silicon dopants, this has the net effect of always aligning the dopant positions between time steps -- leading to a conclusion that the dopants never move, which is known to be false. 
A second, more sophisticated alternative is to use the iterative closest points (ICP) algorithm on the extracted atom positions (a similar technique as used by \citet{roccapriore21identification}). This allows the algorithm to equally weigh the dopant and non-dopant atoms, simply aligning the lattices together. This method leads to acceptable alignments in most cases, but we found it to be quite sensitive to failures in atom detection.

To address these issues, we instead use a denoising convolutional neural network \citep{jain08natural} that solves the alignment problem directly from scans of the system. Given a stack of image observations $(o_1, ..., o_n)$, the network is trained to predict the drift $d_n$ between $o_{n-1}$ and $o_{n}$, as a two-dimensional vector. Historical observations $o_1, \ldots, o_{n-2}$ serve to provide context and noise reduction, but are not used in the loss; as they have already been approximately aligned with $o_{n-1}$, they provide additional information about the needed shift. After the network is applied, we take its prediction $\hat{d}_n$ and shift the current observation by it to align it to the prior observations, and add it to the array of historical observations.

To train this network, we generated a dataset of synthetic trajectories, each consisting of sequential image observations of a doped graphene system under random, correlated drift. To add robustness, we applied both trajectory-wide augmentation by randomly dropping atoms, introducing regions of bright contamination, and adding synthetic large holes to the system. We then simulated drift, treating the direction and magnitude of the drift as a temporally correlated random variables. Finally, we took a series of synthetic scans with the simulated cumulative drifts applied; for each scan, we also randomly perturbed the system, occasionally dropping or moving atoms.
We parameterized the network as six convolutional layers followed by downsampling, followed by a single fully-connected layer. We trained drift correction prediction with mean squared error.

Note that the estimation of ${d}_n$ is a denoising task. Given this, we can iteratively apply our network for more precise drift correction. We find that this has a very large impact on performance. Qualitatively it is clear that the alignment quality increases with additional iterations (see bottom-right panel of \cref{fig:methodology}). Quantitatively, a transition model trained on only single-step-aligned data produces markedly different estimates of the optimal beam position, which are not able to successfully cause transitions when applied to the greedy controller~(see ``Past Neighbor" in~\cref{fig:controller-results-and-controllers}).

\subsection{Data filtering, augmentation, and structure}
\label{sec:dataFiltering}
As there are many possible sources of noise and error in our data, we apply several types of filtering prior to training. Out of a starting data set size of $6,754$ examples, we discard transitions where: {\bf (1)} There is no recorded beam position ($793$ {examples}); {\bf (2)}
 There is not exactly one (1) detected dopant atom before and after the transition ($4$ examples); {\bf (3)} The dopant does not have the expected number of neighbors (3) before and after the transition, either because they are absent, indicating non-pristine graphene, or were not detected ($3,593$ {examples}); and {\bf (4)} the neighbors after the transition did not approximately align with the neighbors before the transition (chosen as an average distance $>$ 0.71\r{A}, $411$ {examples}).
 
These led us to discard approximately 80\% of the data we received from the microscope, resulting in a final dataset of $1,953$ {examples}.
Once filtering has been done, we further post-process our observations to create a uniformized training set. To express beam positions in a consistent format, we translated them to a frame-of-reference relative to the current position of the dopant to be moved, with this atom at the origin. We then label the neighbor closest to the beam position as neighbor 1; we rotate the system, including the beam, such that this atom lies on the x-axis. The neighboring atoms are then numbered accordingly in counter-clockwise order; these are the indices we predict in our classification. We denote ``no movement'' as index 0 in the classification task.
This labeling is illustrated in the top-center panel of \cref{fig:problemDescription}.

In the structure noted above, there is no systematic difference between the three neighboring atoms; only their distance from the beam separates them. As we expect our system to be invariant to both rotation and reflection, we enforce this by adding data augmentation. 
To do this, we first reflect across the x-axis with 50\% probability. This exchanges the second and third neighbors. We then apply 0\degree, 120\degree or 240\degree rotations with equal probability, rotating the neighbor indices accordingly. This leads to an effective sixfold increase in data coverage. 
The top-right panel of \cref{fig:methodology} visualizes the beam positions post-augmentation.

\subsection{Learning the transition dynamics}
\label{sec:learning}
    Experimentally, we modeled our problem as a classification task, estimating $P(S' | s_0, a)$, where $S'$ is the position of the silicon at the next time step, $s_0$ is its current position, and $a$ is the beam dwell action chosen by the user, specified as a two-dimensional coordinate $x$ and a duration in seconds $\Delta t$. To constrain our predictions to respect the observed physical reality (i.e., that the probability of moving to another state should be monotonically increasing in $\Delta t$), we further formulate the problem as predicting transition \textit{rates}~\citep{voter2007introduction}. We decompose this as a total rate $\lambda$ and a categorical distribution over possible next states $y$; this corresponds to decomposing $P(S' | a)$ into a probability of some transition occurring, $P(S' \neq s_0 | s_0, a)$, and a distribution over which next state is chosen if a transition occurs, $P(S' | s_0, a, S' \neq s_0)$. We parameterize $P(S' \neq s_0 | s_0, a)$ as $1 - e^{- \lambda \Delta t}$ -- i.e., as an exponential cumulative distribution function. If desired, per-neighbor rates are simply given by $\lambda y$. 
    We then formulate maximum likelihood loss functions for the total rate and distribution over neighbors:

\begin{align}
    \mathbf{J}_{rate} & = -\mathbb{I}_{(S' \neq s_0)} \cdot (-\lambda \Delta t) - \mathbb{I}_{(S' = s_0)} \log (1 - \exp (-\lambda \Delta t) \\
    \mathbf{J}_{next} & = -\mathbb{I}_{(S' \neq s_0)} (S' \cdot \log y) \\
    \mathbf{J}_{total} & = \mathbf{J}_{rate} + \mathbf{J}_{next} \, .
\end{align}

Given the above loss function, we trained a three-layer neural network using Adam with weight decay~\citep{kingma15adam} and ReLU hidden layer nonlinearities, trained with the cross-entropy loss for 500 epochs using batch size 256. We predicted $\lambda$ with a softplus activation and $y$ with a softmax. We normalize all inputs to the network prior to training, for stability.

To improve the robustness of our transition model to initializations, we trained an ensemble of transition predictors of bootstrap-resampled datasets.
In addition to giving us the ability to esimate uncertainties, this significantly improved our overall accuracy and robustness 
To support more rapid inference, we also distilled this ensemble to a single transition predictor. By using widely-sampled random beam positions and a large number of training steps, this distillation could be made to fairly precisely match the ensemble predictions.

We display an example of the probability contours found by our system in the center panel of \cref{fig:problemDescription}. Note that we are overlaying three different probability distributions (one for each neighbour), where the colours are used to distinguish them. Our main finding confirms what was anecdotally held to be true by the microscopy community:

\begin{tcolorbox}[leftrule=1.5mm,top=1mm,bottom=0mm]
To induce the dopant to transition to one of its neighbours, the optimal beam placement is directly on the neighbour, with a 50.0\% (95\% CI 0.246 to 0.754) probability of causing a transition with a five second dwell time, using a nominal beam current of 90 pA.
\end{tcolorbox}

Perhaps more relevant than dwell time is the number of electrons emitted, which is a function of both dwell time and the beam current. For our experiments we used a beam current of 90 pA, resulting in approximately 3 billion electrons in a five second period.

\section{Empirical evaluation}

\begin{figure}[!t]
\begin{tabular}{c | c}
    \begin{subfigure}{0.525\textwidth}
        \includegraphics[width=\textwidth]{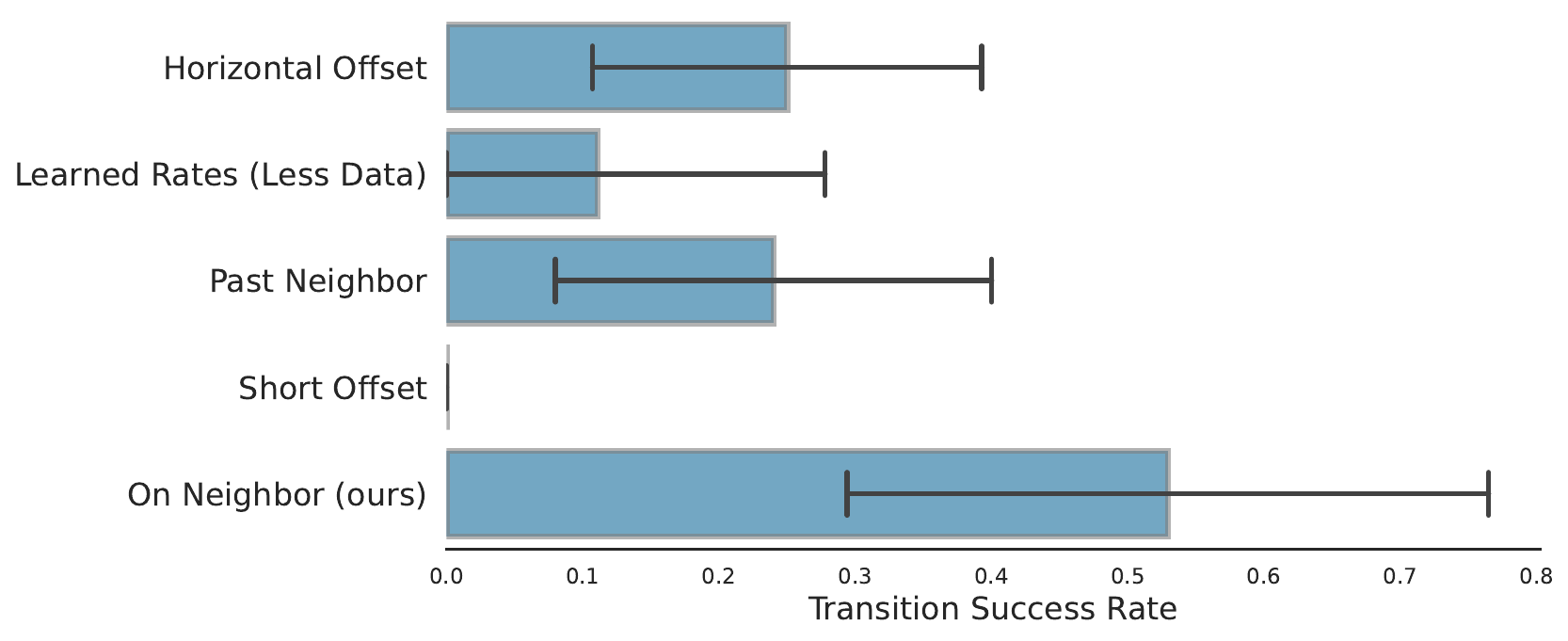}
    \end{subfigure} &
    \begin{subfigure}{0.375\textwidth}
        \includegraphics[width=\textwidth]{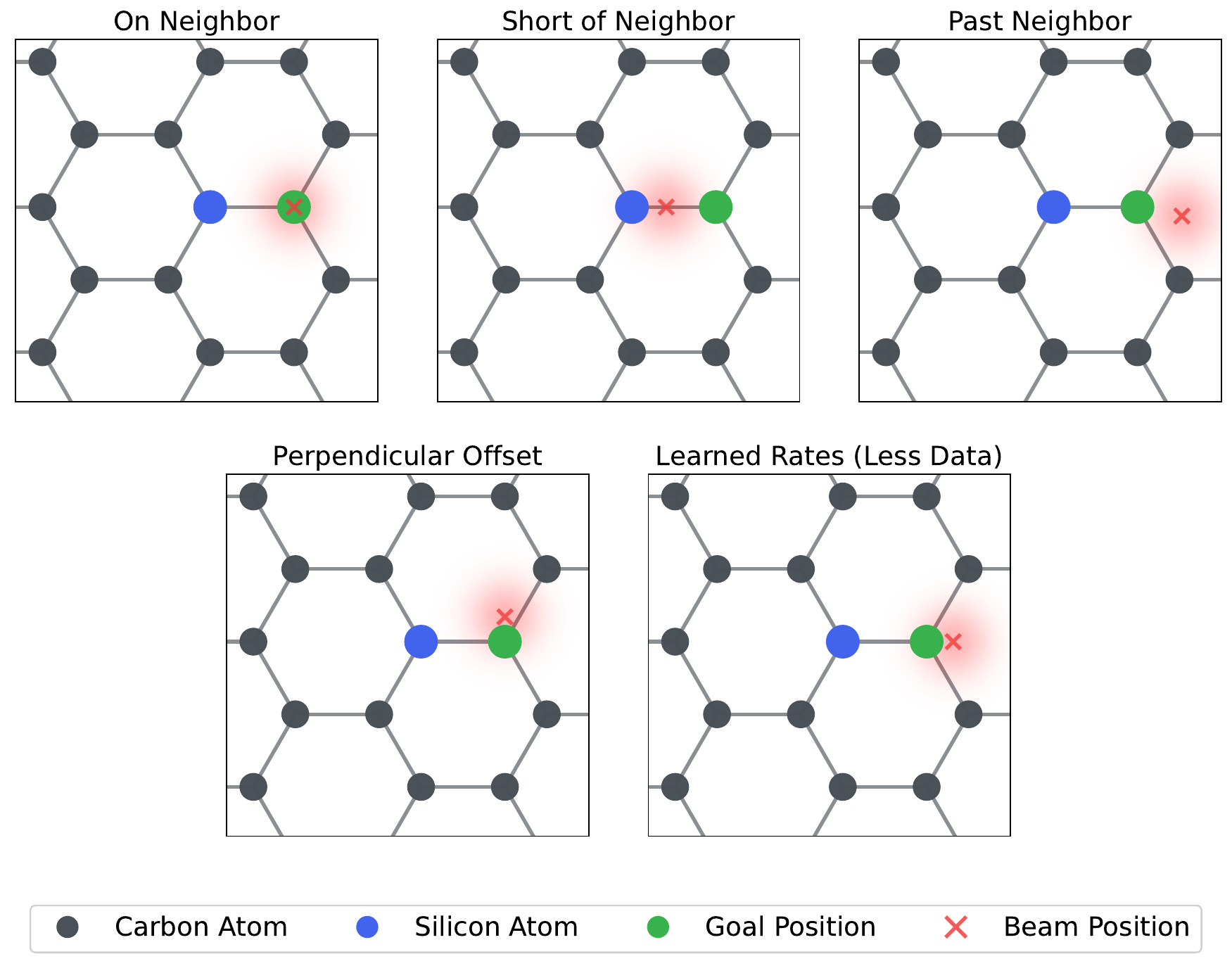}
    \end{subfigure}
\end{tabular}
\caption{\textbf{Left:} The proportion of transitions that induced the intended transition in under a $5$ second dwell. \textbf{Right:} The beam placement strategies considered in our experiments.}\label{fig:controller-results-and-controllers}
\end{figure}

While our main finding is consistent with previously held beliefs in the community, we do not have ground-truth data for the learned transition probabilities to quantitatively assess the accuracy of our predictions.
However,
as discussed in \cref{sec:problemDescription}, the purpose of learning these transition probabilities is to be able to automate atomic manipulation, so in this section we evaluate the efficacy of our learned transition functions for this purpose. Specifically, our experiments are conducted as follows:
\begin{enumerate}
    \item Start from a configuration with one dopant and 3-fold connections to neighbours, with the field-of-view (FOV) centered at the dopant.
    \item Pick an arbitrary carbon atom as the goal position.
    \item Focus the beam on a position dictated by one of the strategies (defined below).
    \item Acquire an image to determine if we caused a transition.
    \item Repeat above steps until the dopant has arrived at the goal position, or until we have reached the maximum allowable attempts.
\end{enumerate}

In order to determine that on-neighbour is in fact the optimal beam placement, we use a few different strategies for beam placement, detailed below and illustrated in Figure~\ref{fig:controller-results-and-controllers} (Right).
\begin{itemize}
    \item {\bf On neighbour:} our proposed optimal strategy.
    \item {\bf Short of neighbour:} place the beam in between the dopant and its neighbour.
    \item {\bf Past neighbour:} place the beam beyond the neighbour atom
    \item {\bf Perpendicular offset:} offset the beam perpendicularly from neighbour.
    \item {\bf Learned dynamics (less data):} we ran the learning method detailed in \cref{sec:learning}, but with approximately half of the collected data. When doing so, the resulting "optimal" beam placement was just past the neighbour.
\end{itemize}

The first controller was chosen as the optimal policy according to our most recent learned transition models; the second, third, and fourth were chosen to test how deviations from this policy affected results, in a systematic way. The fifth and final controller represents the optimal policy according to an early iteration of our learned models, and serves to validate the progress made as additional data was introduced.

For consistency, we used a constant $5$s dwell time at a nominal beam current of 90 pA for all agents. We measured the number of times the electron beam was used to try and induce a dopant transition and report the findings in Figure~\ref{fig:controller-results-and-controllers}.
We observe that  the on neighbor strategy induces the intended transition on average approximately $50\%$ of the time, whereas the other approaches are below $25\%$.

\begin{figure}
\includegraphics[width=\textwidth]{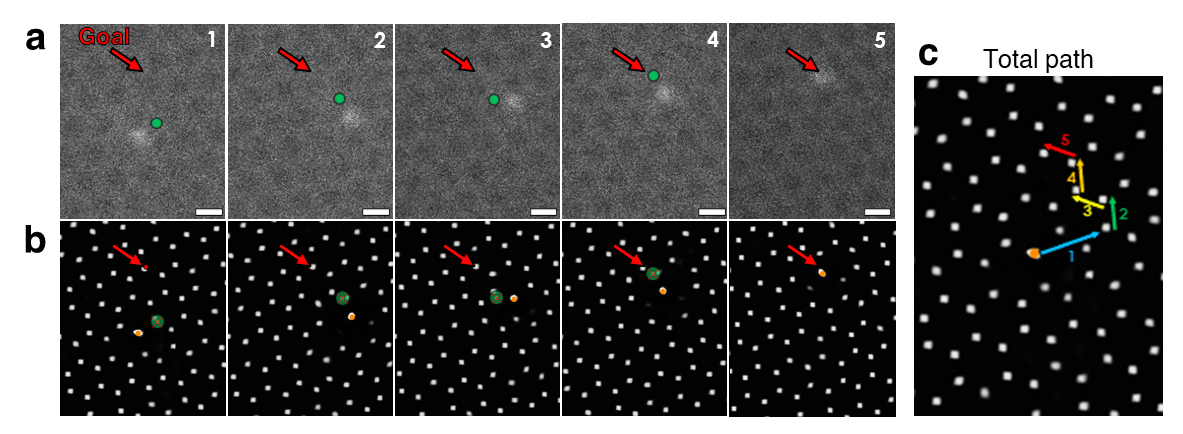}
\caption{Experimental realization of automation of silicon motion throughout graphene using learned transition rates. HAADF-STEM image snapshots of five steps are shown in {\bf a}, while corresponding predicted images are shown below in {\bf b}. Images are aligned to the initially selected goal position, denoted by red arrows, and beam positions shown in green circles. The silicon is the bright atom in {\bf a} due to its higher atomic number, and is shown by an orange circle in the predicted images. The path summary of the silicon atom between each step is shown in {\bf c}. Scale bars all 2 Angstroms.}\label{fig:experimental_deploy_example}
\end{figure}

Automated silicon manipulation was experimentally realized on a Nion UltraSTEM100 microscope, and an example using the on neighbor protocols is shown in {\bf Figure~\ref{fig:experimental_deploy_example}}. Here, a goal location - shown by red arrows in {\bf Figure~\ref{fig:experimental_deploy_example}a} - is randomly chosen relative to the silicon, and the controller directs the electron beam on carbon neighbors to drive the silicon to the goal position. To accomplish this task automatically, the atomic coordinates must be known in as close to real time as possible; here, deep ensemble neural networks were used which have been shown to be both fast and robust even to out of distribution effects \citep{Ghosh2021,Roccapriore2022}. These predictions are shown in \textbf{Figure~\ref{fig:experimental_deploy_example}b} where the silicon is depicted by a larger orange circle. Fast access to the atom positions allows the chosen controller to then place the beam at appropriate positions relative to the silicon to fully automate this process. Note that the first step in Figure \ref{fig:experimental_deploy_example} has caused the silicon to jump multiple sites, whereas the remaining steps cause single atomic jumps. Real-time detector feedback was not used to indicate if the silicon transitioned earlier than the designated dwell time, and therefore this was one limitation in the current work, which can explain the (albeit lower probability) longer initial jump. We also note that this behavior can be exploited for more efficient beam manipulation; i.e., one beam position can intentionally induce more than one atomic jump.

\section{In-depth empirical analyses}

\subsection{Synthetic data}
To test the learning behavior of our algorithm beyond our relatively small real-world dataset, we generated many datasets of simulated microscope interactions. We use these datasets to evaluate the scaling behavior of our algorithm, demonstrate that models ultimately converge to accurate estimates with sufficient data, and to evaluate our algorithmic and hyperparameter decisions.

\begin{figure}
\begin{tabular}{c | c}
    \begin{subfigure}[t]{0.60\textwidth}
        \vskip 0.5pt
        \includegraphics[width=0.95\textwidth]{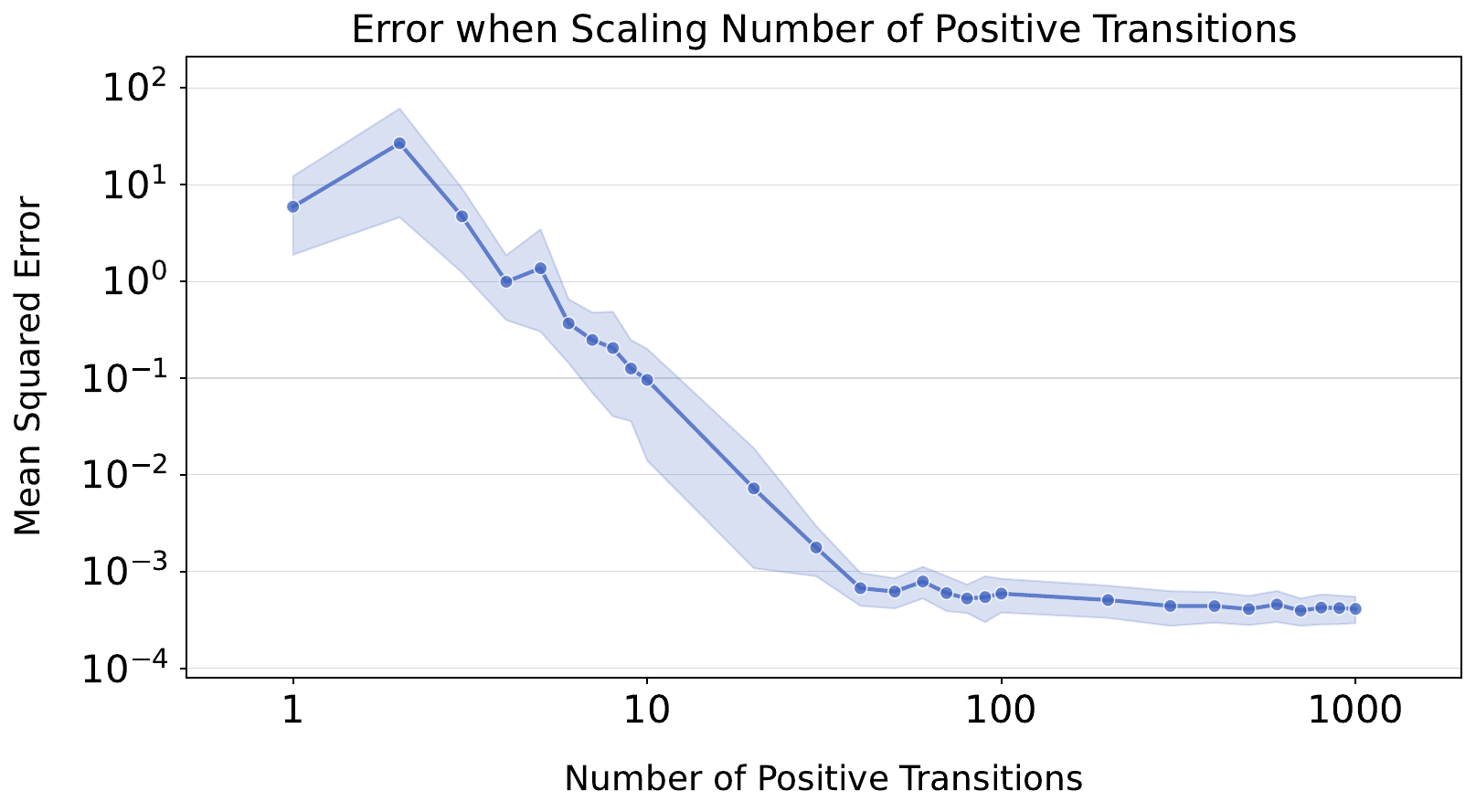}
    \end{subfigure} &
    \begin{subfigure}[t]{0.3\textwidth}
        \vskip 0pt
        \centering
        \makebox[\linewidth][c]{\hspace{8mm}{\fontsize{7pt}{14pt}\textsf{Synthetic Rate Functions}}}
        \begin{tabular}{m{0.48\linewidth}m{0.48\linewidth}}
            {\includegraphics[width=0.5\textwidth]{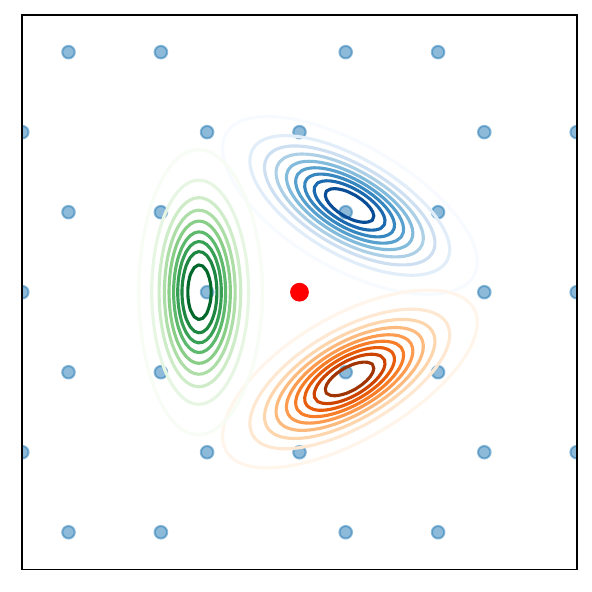}} &
            {\includegraphics[width=0.5\textwidth]{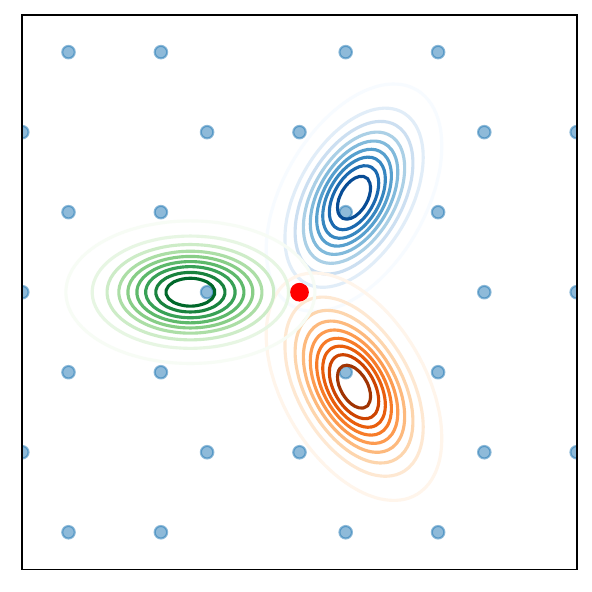}} \\
            {\includegraphics[width=0.5\textwidth]{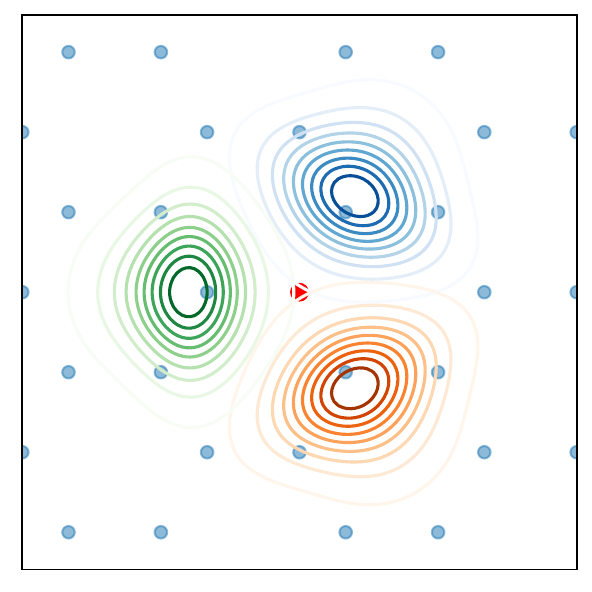}} &
            {\includegraphics[width=0.5\textwidth]{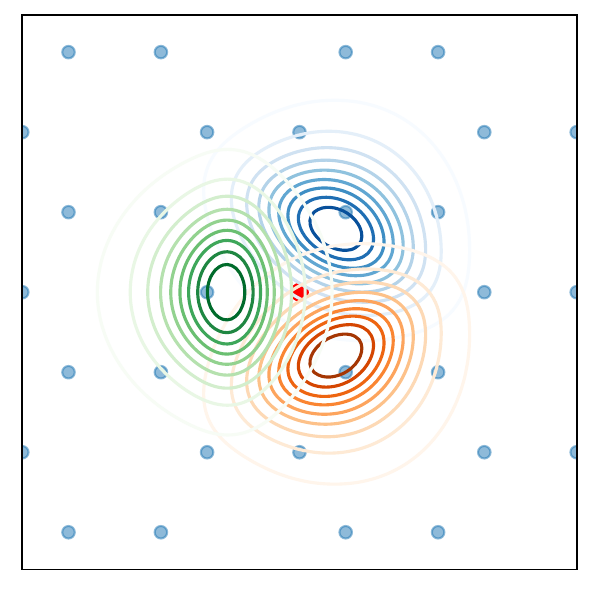}}
        \end{tabular}
    \end{subfigure}
\end{tabular}
\caption{{\bf Left: } Number of positive transitions in dataset versus the sum of squared prediction error. {\bf Right: } A visualization of some of the synthetic rate functions overlayed on a sheet of graphene. The contours represent the rate of transitioning to the associated neighbor.}\label{fig:scaling-and-synthetic-rates}
\end{figure}

To generate these datasets, we sampled synthetic transition probability distributions as mixtures-of-Gaussians,\footnote{Chosen because they generalize the family of rate functions we expect to see in the real world by permitting multimodality, with the Gaussian nature of the distributions being sensible as the electron beam is approximately a Gaussian beam.} each giving the non-normalized transition rates to neighboring states given a certain beam position (equivalent to the predicted per-state rates $\lambda y$ in our neural network model). Figure~\ref{fig:scaling-and-synthetic-rates} (Right) is representative of the type of synthetic rate functions used throughout this section. Doing so grants us ground-truth data for transition probabilities, allowing us to quantitatively assess the accuracy of our learned transition model.

When creating a dataset, we first generate a simulated graphene sheet with a single dopant. We then uniformly sample random actions within 2\r{A} of the silicon and simulate the transitions of the silicon according to the synthetic rate function, continuing until a certain number of positive transitions have been observed; as positive transitions are generally far rarer than negative transitions, they are the critical determiner of effective dataset size. To simplify comparisons between datasets, we enforce that the synthetic rate functions for each dataset have the same maximum value, preventing us from sampling datasets that are entirely positive transitions or almost entirely negative transitions.

We can then use a synthetic dataset to evaluate a learning algorithm by training on it and directly comparing its predictions (here, the predicted rates $\lambda y$) to the known true values across a large grid of beam positions (a uniformly-spaced 2d grid surrounding the dopant). This provides a measure of the total accuracy of the learned rate function across a full range of possible beam locations, even those that would not normally be chosen when attempting to induce a transition.

\subsection{Data scaling}
To test the data scaling of our model -- and show that it converges to near-perfect predictions in the limit of large datasets -- we use this evaluation procedure at a range of data scales. We start by collecting 30 datasets with at least 1,000 positive transitions each.
We then train our models on a range of subsampled scales from these datasets, and report a bootstrapped confidence intervals over our our 30 synthetic datasets at each scale in Figure~\ref{fig:scaling-and-synthetic-rates} (Left). As expected, we find that our model has very poor performance with limited amounts of data, but rapidly improves once hundreds of positive transitions are available, matching our own experience of the model's improvement as additional data was collected.

To evaluate the utility of these learned models for control, we measure the \textit{regret}, or total suboptimality caused by acting according to our learned model compared to the true rate function. To do this, we take the \texttt{argmax} of a learned model $\hat{R}$, and evaluate the true synthetic rate function $R$ at this point. We then define the regret as the difference between the true maximum of the rate function and this value:
\begin{align}
\text{Regret}(\hat{R}) \triangleq \max_a R(a) - R(\arg \max_a \hat{R}(a))
\end{align}
Intuitively, this measure captures how much slower control according to a learned rate function would be than control according to the true rate function. 

\begin{figure}
\centering
\begin{tabular}{c | c}
    \begin{subfigure}{0.475\textwidth}
        \includegraphics[width=\textwidth]{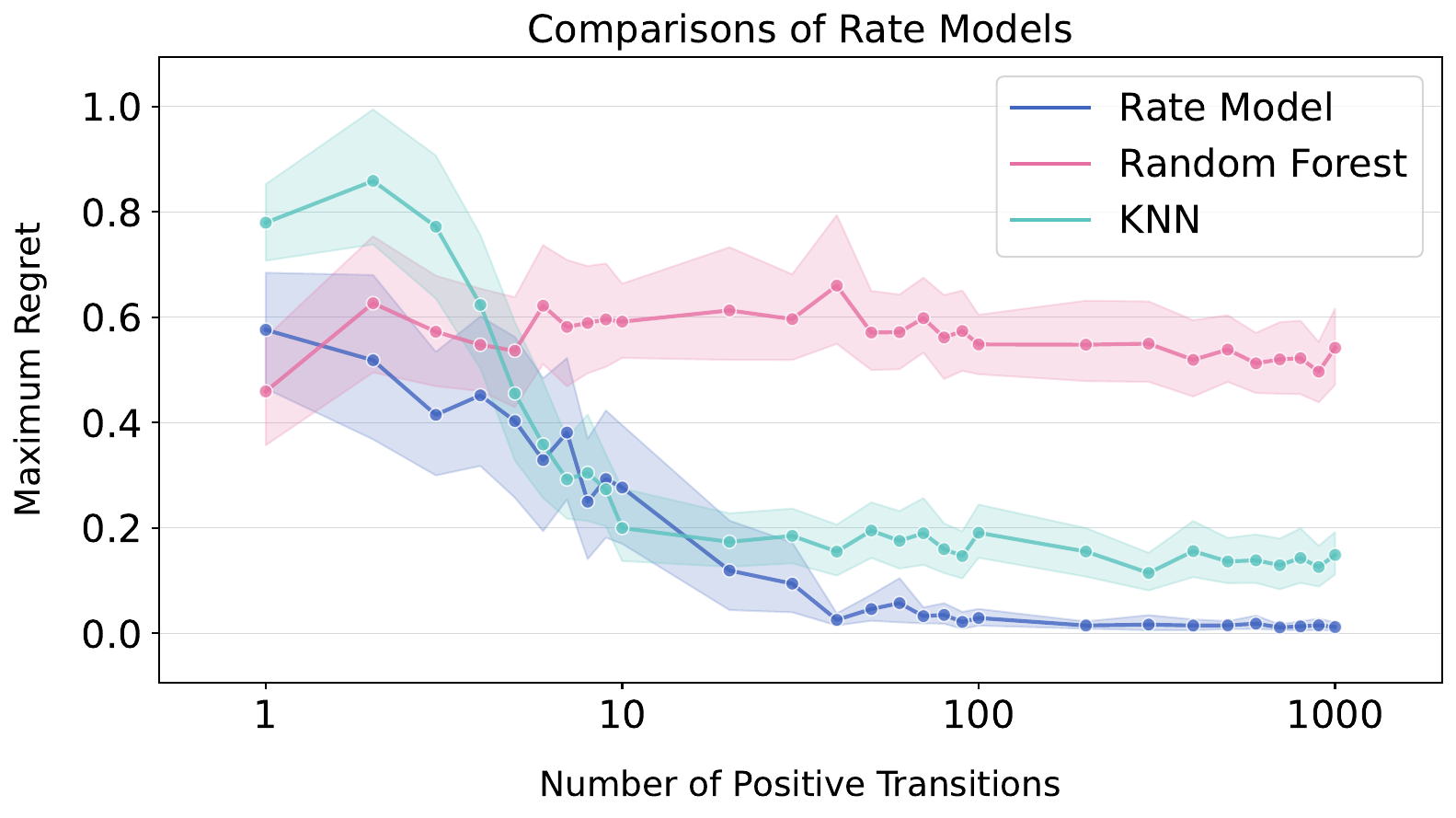}
    \end{subfigure}&%
    \begin{subfigure}{0.475\textwidth}
        \includegraphics[width=\textwidth]{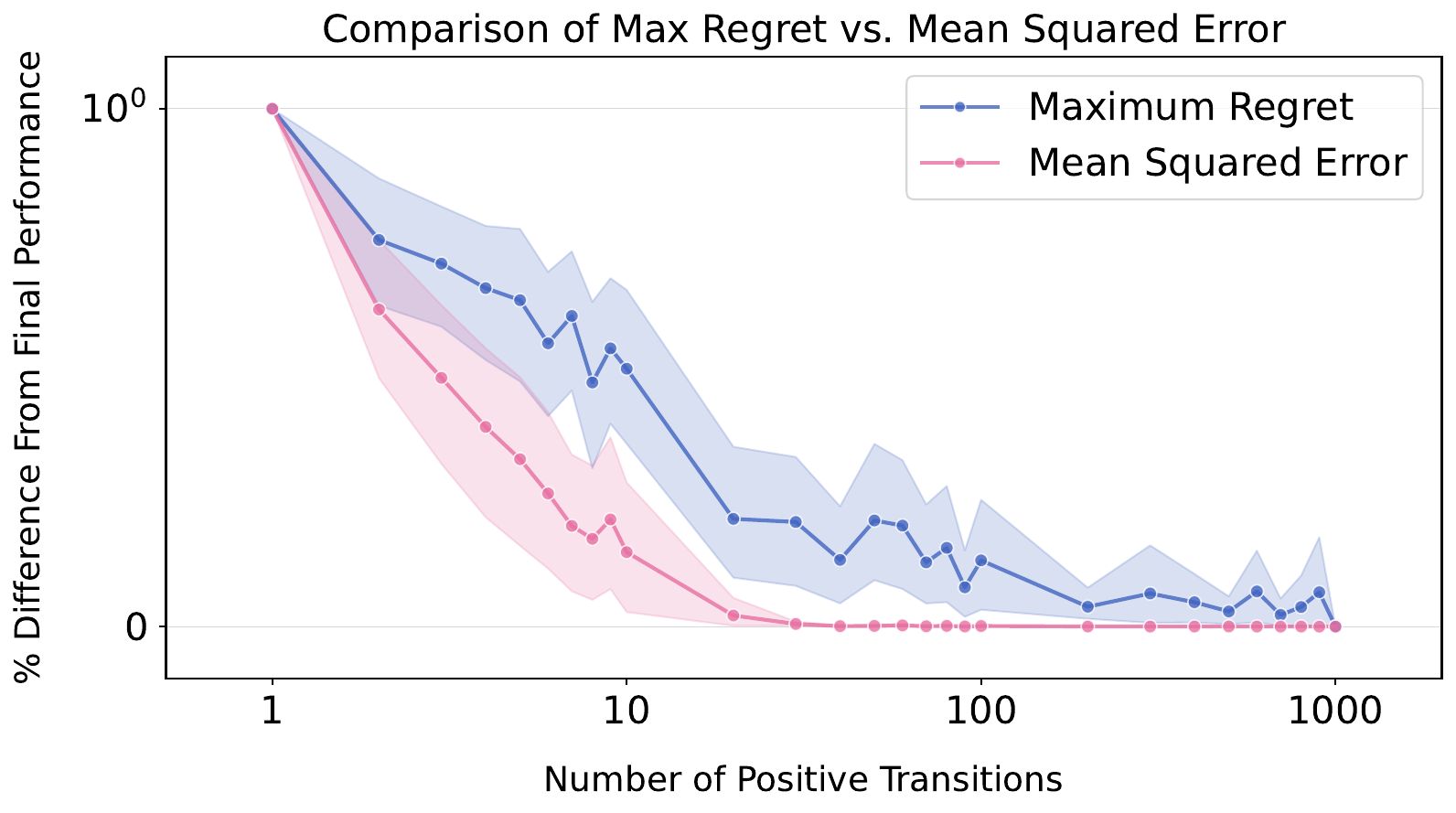}
    \end{subfigure}
\end{tabular}
\caption{{\bf Left: } Positive transitions in dataset versus the incurred regret for our rate model versus a $k$-nearest neighbors and a random forest classifier. {\bf Right: } Comparing the rate of convergence when measuring maximum regret versus mean squared error.}\label{fig:baselines_and_regret}
\end{figure}

Figure~\ref{fig:baselines_and_regret} (Right) depicts the relative rate of convergence between the mean-square error and the regret of the true rate function. We find that the model tends to converge more quickly to the global shape of the rate function (as measured by the mean squared error) as compared to true \texttt{argmax} (as measured by the regret).
This matches our empirical experience where the learned rate functions can look similar although their \texttt{argmax} can differ (e.g., Learned Rates (Less Data) vs Learned Rates in Figure~\ref{fig:controller-results-and-controllers} (Right)).

\subsection{Hyperparameter and algorithmic evaluation}
\begin{figure}
\centering
\begin{tabular}{c | c}
    \begin{subfigure}[t]{0.65\textwidth}
        \vskip 16pt
        \includegraphics[width=\textwidth]{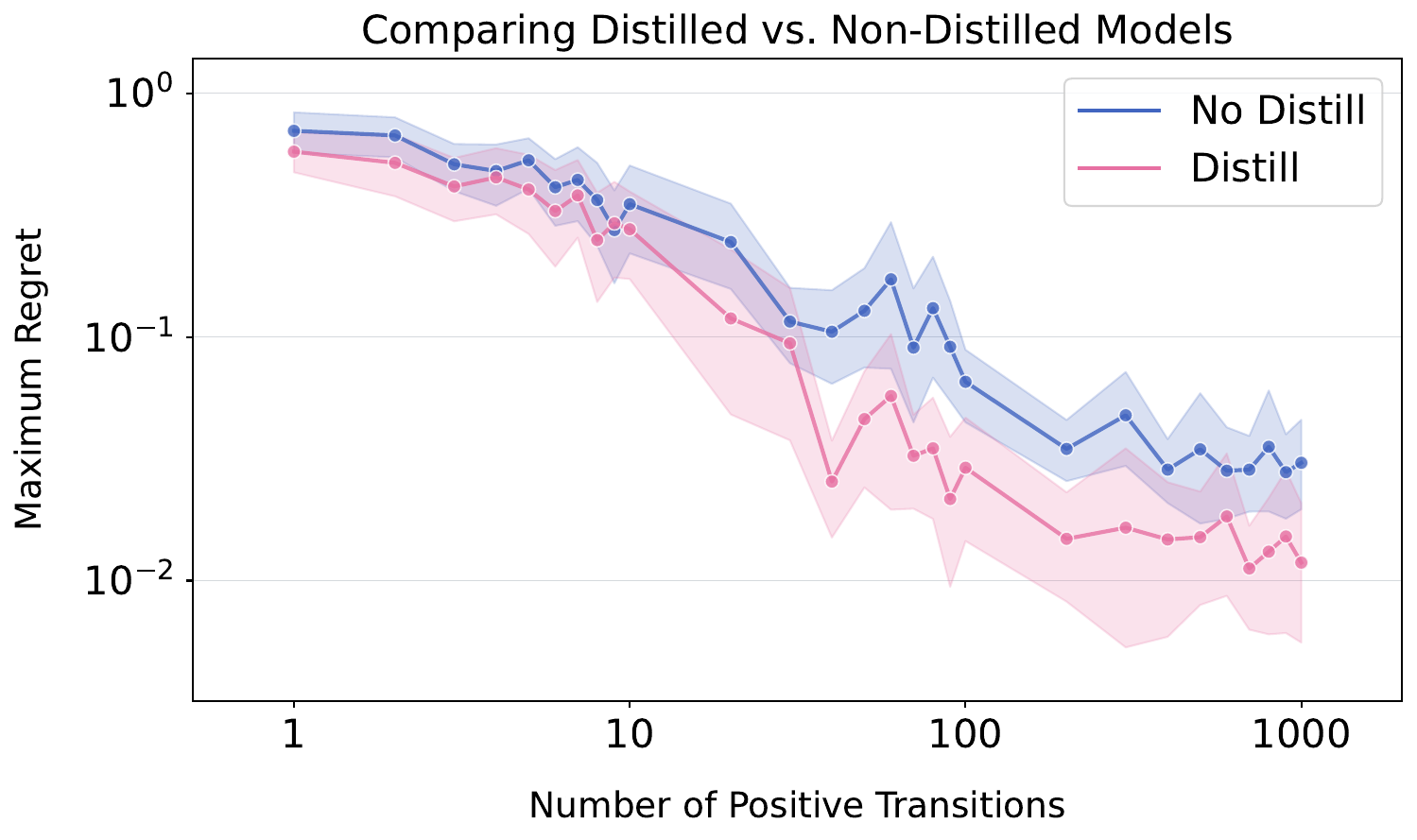}
    \end{subfigure} &
    \begin{subfigure}[t]{0.25\textwidth}
        \vskip 0pt
        \begin{tabular}[t]{m{\linewidth}}
            \makebox[0.89\linewidth][c]{{{\fontsize{6pt}{14pt}\textsf{Distilled Rate Functions}}}}
            \vskip 0.5pt
            \includegraphics[width=0.85\linewidth]{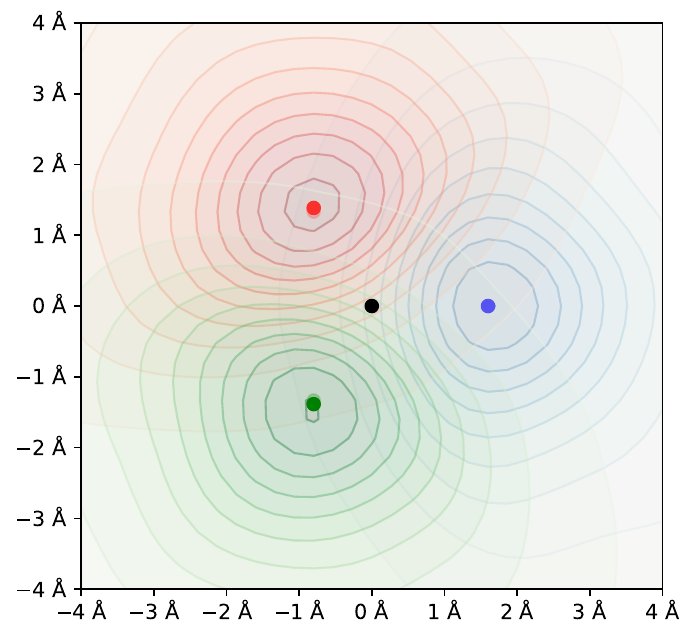} \\
            \makebox[0.89\linewidth][c]{{{\fontsize{6pt}{14pt}\textsf{Non-Distilled Rate Functions}}}}
            \vskip 0.5pt
            \includegraphics[width=0.85\linewidth]{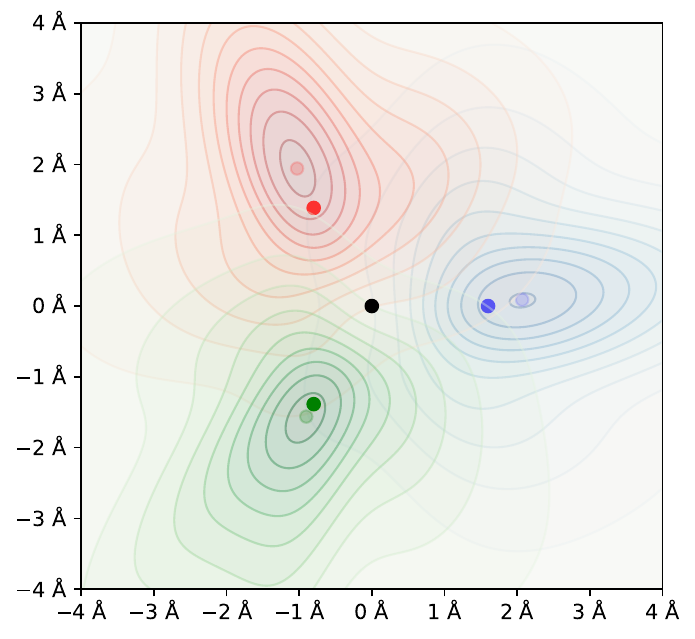}
        \end{tabular}
    \end{subfigure}
\end{tabular}
\caption{Comparisons of transition prediction models trained a single models (top) and as a distilled ensemble of multiple models (bottom) on a suite of synthetic datasets (left) and on real data (right). Note that the distilled ensemble has both better overall regret on synthetic data and shows more regular predicted rates on real data.}\label{fig:distillation}
\end{figure}

Beyond scaling experiments, we leverage this synthetic evaluation framework to validate several hyperparameter decisions that we found to be important in the real data setting. Firstly, in Figure~\ref{fig:distillation} we examine the importance of training an ensemble of transition prediction models and then distilling them. We find that distillation significantly improves regret at all data scales on synthetic data; we visualize examples of distilled and non-distilled models trained on real data, to illustrate how this might take place. We find that the result is largest in the intermediate-data regime when meaningful transition models can be learned but inaccuracies are common, matching our experience in the progression of real data experiments.

We also evaluate other possible machine learning paradigms, including families of classical algorithms such as K-nearest-neighbors and random forests in Figure~\ref{fig:baselines_and_regret}~(Left). We find that these algorithms are also capable of learning reasonable transition models, but generally require significantly more data to do so and ultimately converge to worse solutions, suggesting that the inductive biases provided by modern neural networks are important to this task.

\section{Conclusion}
 
The last note left by Richard Feynman stated “What I cannot create, I do not understand.” Building solid state quantum computers, creating nano-robots, and designing new classes of biological molecules and catalysts alike requires the capability to manipulate and assemble matter atom by atom, probe the resulting structures, and connect them to the macroscopic world; all this necessitates accurate estimates of the transition dynamics induced by sub-atomically focused electron beams. Until now, the elements of relevant knowledge have been limited
to a few research groups, and atomic manipulation has been performed via direct control by
human operator one beam positioning at a time. The characteristic timescale of human-operated
experiments vastly exceeds the intrinsic latency of the electron microscope, for
which hundreds of fabrications steps per second should be possible. Similarly, human control
necessarily lacks precision, reproducibility, and systematic error correction capabilities. While
sufficient for a proof of concept, atomic scale fabrication with the precision and throughput
necessary for applications such as nanopore fabrication for protein sequencing, molecule screening platforms for physics and biology, and particularly quantum communication, sensing,
and computing devices requires moving beyond the current human control paradigm.

Our work is a robust first step for determining transition probabilities via machine learning, and paves the way for further advances in this space. The scenario we considered in this work is somewhat idealized: we limit ourselves to single dopant and 3-way lattices for our learned dynamics. Nevertheless, these settings allowed us to confirm, via a {\em data-driven approach}, the commonly held belief that placing the electron beam directly on the neighbour has the highest probability of inducing a transition of the dopant. Going forward, we will be exploring broader settings: multiple dopants, graphene with 4-way connections and aberrations (such as holes).

\backmatter

\bmhead{Supplementary information}

The code we used for this work is publicly available at \href{https://github.com/google/putting-dune}{https://github.com/google/putting-dune}.

\bmhead{Acknowledgments}

Autonomous STEM research was supported by the Center for Nanophase Materials Sciences (CNMS) proposal [CNMS2022-B-01647], which is a US Department of Energy, Office of Science User Facility at Oak Ridge National Laboratory. The authors would also like to thank their colleagues at Google DeepMind for their useful feedback on this work.

\newpage
 
\bibliography{rateLearning}

\end{document}